\documentclass[letter]{aa}     % for the letters
\usepackage{graphicx}
\usepackage{txfonts}
\usepackage{subcaption}         % necessary for continued figures, example in section 3
\usepackage{lscape}             % to rotate a single page table, example in appendix.
\usepackage{placeins}           % useful with \FloatBarrier, to keep 
\usepackage{hyperref}
\hypersetup{
    colorlinks=true,
    linkcolor=blue,
    filecolor=blue,
    urlcolor=blue,
    citecolor=blue
}
\begin{document}

%%%%%%%%%%%%%%%%%%%%%%%%%%%%%%%%%%%%%%%%
% if you use custom commands in your title,
% ensure to check your title when submitting!
%%%%%%%%%%%%%%%%%%%%%%%%%%%%%%%%%%%%%%%%
   \title{A solution to the paradox of youth in the Galactic centre}

   % \subtitle{Subtitle}

%%%%%%%%%%%%%%%%%%%%%%%%%%%%%%%%%%%%%%%%
% Please separate each author with the \and command
%
% Use the \corrauth to provide the corresponding
% author address. It will be automatically inserted as 
% footnote in the PDF output.
%
% Please DO NOT include ORCIDs next to author names.
% Instead, please provide an active address for each coauthor:
% it will be automatically extracted by EDPS editorial system, 
% and co-authors will be be able to authenticate their ORCID.
%
% Only authenticated ORCIDs will be taken into account.
% ORCIDs included here will be removed.
%%%%%%%%%%%%%%%%%%%%%%%%%%%%%%%%%%%%%%%%

   \author{Rodrigo P. Silva \inst{1,2,3}
          \and
          Alexandre C. M. Correia\inst{1,4}
          \and
          Tjarda C. N. Boekholt \inst{5}
          \and
          Paulo J. V. Garcia \inst{2,3}
          }

   \institute{
        CFisUC, Departamento de F\'isica, Universidade de Coimbra, 3004-516 Coimbra, Portugal 
        \and
        Faculdade de Engenharia, Universidade do Porto, Rua Dr. Roberto Frias, 4200-465 Porto, Portugal
        \and
        CENTRA - Centro de Astrof\'isica e Gravita\c c\~ao, IST, Universidade de Lisboa, 1049-001 Lisboa, Portugal
        \and
        LTE, Observatoire de Paris, Universit\'e PSL, Sorbonne Universit\'e, CNRS, 75014 Paris, France
        \and
        Anton Pannekoek Institute for Astronomy, University of Amsterdam, NL-1090 GE Amsterdam, The Netherlands
        }

   \date{}

% \abstract{}{}{}{}{}
% 5 {} token are mandatory
 
  \abstract{The supermassive black hole at the Galactic centre, Sgr A*, is surrounded by an S cluster of young stars. These stars are predominantly B-type stars, but their origin remains uncertain. Unlike massive stars in the Galactic field, which exhibit a binary fraction of about $70\%$, those in the S cluster show a significantly lower fraction. It is key to understand this discrepancy for constraining their formation. We show that the binary deficit arises naturally from dynamical interactions between stellar binaries and Sgr A* in an in situ formation scenario. Using K-amplitude data, we derived an observational binary fraction of $f_{\mathrm{obs}} = 0.43 \pm 0.09$ for the S-cluster stars. Our dynamical modelling predicts that interactions with Sgr A* cause 18\% of the binaries to merge, 20\% to be disrupted, and 62\% to survive, yielding a theoretical binary fraction of $f_{\mathrm{theo}} = 0.38 \pm 0.10$. The predicted binary fraction with distance from Sgr A* is also consistent with independent observational constraints. These results show that the decline in binarity towards the Galactic centre is driven by stronger tidal forces, and they support an in situ origin of the S cluster in the same event that produced the Fermi bubbles.}

   \keywords{Celestial mechanics -- Stars: formation -- Stars: binaries: general -- Stars: kinematics and dynamics --  Galaxy: centre}

   \maketitle
\nolinenumbers

%%%%%%%%%%%%%%%%%%%%%%%%%%%%%%%%%%%%%%%%%%%%%%%%%%%%%%%%%%%%%%
\section{Introduction}
The S cluster is a group of stars located within the central $\sim 0.04$~pc of the Milky Way, orbiting the supermassive black hole Sgr~A*  \citep[e.g.][]{2008ApJ...689.1044G, 2009ApJ...692.1075G, 2010RvMP...82.3121G}. The monitoring of these S stars has provided key observational evidence of the black hole mass and compactness \citep[e.g.][]{2008ApJ...689.1044G, 2019A&A...625L..10G, 2022A&A...657L..12G}.
The S cluster  is dominated by young B-type stars ($\lesssim 15\,\mathrm{Myr}$, mean age $\sim 6\,\mathrm{Myr}$; \citealt{2005ApJ...628..246E,2017ApJ...847..120H}) with a thermal eccentricity distribution \citep{2017ApJ...837...30G}, isotropic orbital angular momentum vectors \citep{2009ApJ...692.1075G}, and a lower binary fraction ($\lesssim 47\%$; \citealt{2023ApJ...948...94C}) than field stars ($69 \pm 9\%$; \citealt{2012Sci...337..444S}). The surrounding clockwise disk ($0.04 - 0.5$pc) contains B-type, O-type, and Wolf--Rayet stars of a comparable age, but with a unimodal eccentricity distribution and aligned angular momentum vectors \citep{2014ApJ...783..131Y,2017ApJ...847..120H}.

The so-called paradox of youth of the  S stars \citep{2003ApJ...586L.127G} highlights the discrepancy of how young stars like these could have formed in the immediate vicinity of a supermassive black hole. In this environment, the tidal field makes conventional in situ star formation problematic. Molecular clouds are prevented from fragmenting under typical conditions \citep{2012ApJ...749..168M}. Several alternative scenarios have been proposed.

A possible in situ scenario is gas-shell fragmentation driven by a brief Sgr~A* outflow $\sim 6$~Myr ago \citep{2018MNRAS.478L.127N}, motivated in part by the Fermi bubbles \citep{2010ApJ...724.1044S}. Compression and fragmentation of the resulting shell form the S stars \citep{2018MNRAS.478L.127N}, while standard molecular-disk fragmentation forms the clockwise-disk stars \citep{2012ApJ...749..168M}. This naturally explains their coevality and the S-cluster thermal eccentricity and isotropic angular momentum distributions \citep{2018MNRAS.478L.127N}. Its main difficulties are the unclear origin of hypervelocity stars from the Galactic centre \citep[e.g.][]{2020MNRAS.491.2465K,2026A&A...709A.117C}, although rare ejections from bound binaries might help \citep{2026A&A...706A..63S}, and the need for a super-Eddington event plus a cloud with low angular momentum \citep{2018MNRAS.478L.127N}. However, massive galactic outflows may permit star formation \citep{2017Natur.544..202M}, and the required shell mass ($\gtrsim 10^{5}\,M_{\odot}$) is only $\gtrsim 0.5\%$ of the gas reservoir in the central molecular zone ($\sim 2 \times 10^7\,M_{\odot}$; \citealt{2025ApJ...984..156B}), which is consistent with the gas mass needed to form the disk stars \citep{2008Sci...321.1060B, 2009MNRAS.394..191H}.

An alternative ex situ formation scenario is disk migration, in which the S stars form in the clockwise disk and migrate inward \citep{2009ApJ...702..884P}. However, this might require an intermediate-mass black hole, which is now strongly constrained \citep{2023A&A...672A..63G}. It does not explain why migration should be efficient for B stars, but not for O-type and Wolf--Rayet stars \citep{2010RvMP...82.3121G}. Finally, migrated stars are expected to retain modest eccentricities, making the thermal eccentricity distribution difficult to reproduce \citep{2009ApJ...702..884P,2014ApJ...786L..14C}. 
Moreover, binary migration might also leave a non-zero binary fraction if some systems avoid Lidov--Kozai-driven mergers, which might be consistent with the newly identified D9 binary in the S cluster \citep{2024NatCo..1510608P}. However, the processes required to isotropise the S-star orbits would take longer than their stellar ages \citep{2013ApJ...763L..10A}.

Another ex situ formation scenario is the Hills mechanism \citep{1988Natur.331..687H}, in which a binary is tidally disrupted by Sgr~A*, leaving one star bound and the other ejected. However, known high-velocity stars associated with the Galactic centre are much older than the S stars \citep{2020MNRAS.491.2465K,2026A&A...709A.117C}. In addition, this mechanism neither explains the clockwise disk \citep{2013ApJ...764..155L} nor reproduces the S stars thermal eccentricity distribution \citep{2020ApJ...896..137G}. Furthermore, the Hills mechanism naturally places stars on orbits around Sgr~A*, but should leave a virtually zero binary fraction, which is inconsistent with the D9 binary in the S cluster \citep{2024NatCo..1510608P}.

Among these scenarios, gas-shell fragmentation remains the most compelling formation channel for the S stars. An apparent binary-fraction dichotomy has emerged between the S stars and surrounding young populations, however. \citet{2023ApJ...948...94C} reported an upper limit of $47\%$ and \citet{2024ApJ...964..164G} a lower limit of $42\%$. We revisit this issue using updated binary statistics, including the recently identified binary D9 \citep{2024NatCo..1510608P}, together with dynamical simulations of binaries near Sgr~A*. We show that the observations can be reproduced assuming the binary fraction of massive stars in the Galactic field \citep{2012Sci...337..444S}, supporting in situ formation of the S stars in the same event that produced the clockwise disk and launched the Fermi bubbles.

\vspace{-1em}
%%%%%%%%%%%%%%%%%%%%%%%%%%%%%%%%%%%%%%%%%%%%%%%%%%%%%%%%%%%%%%%
\section{Observational limits of the binary fraction}~\label{observational_limits}
 \citet{2023ApJ...948...94C} derived an upper limit on the binary fraction at a distance of $\sim 0.02$~pc from Sgr~A*. 
To obtain this constraint, they generated a synthetic binary population based on the orbital period, $P$, eccentricity, $e$, and mass ratio, $q$, distributions from \citet{2012Sci...337..444S}. 

Following this approach, we generated $10^5$ binaries.
Subsequently, we computed the distribution of the radial velocity semi-amplitude, $K$, using the same formalism as \citep{2023ApJ...948...94C},
\begin{equation} \label{K_amplitude}
    K_{\rm A} = \left(\frac{2\pi G}{P}\right)^{1/3} 
        \frac{m_{\rm B} \sin I}{\left(m_{\rm A} + m_{\rm B}\right)^{2/3}} \ ,
\end{equation}
where $G$ is the gravitational constant, $m_{\rm A}$ and $m_{\rm B}$ are the primary and secondary stellar masses, respectively, and $I$ is the inclination of the binary orbital plane relative to the plane of the sky.

This procedure yielded a distribution of $K$ amplitudes corresponding to a population with a binary fraction of 100\%. 
We then varied the binary fraction by randomly replacing a subset of binaries with $K = 0~\mathrm{km~s^{-1}}$, representing single stars. 
For example, a binary fraction of 90\% corresponds to 10,000 single stars with $K = 0~\mathrm{km~s^{-1}}$ and 90,000 binaries with $K$ amplitudes computed from Eq.~\eqref{K_amplitude}. The binary fraction was systematically varied between 10\% and 100\%.

 \citet{2023ApJ...948...94C} compared each $K$ amplitude drawn from the simulated distribution with the $K$ limits of the 16 early-type stars listed in their Table~7. For each trial, the number of detections might range from 0 (if all sampled values fell below the observational $K$ limits) to 16 (if all exceed them). This process was repeated $10^5$ times for each assumed binary fraction.

Our approach differed in two key aspects. 
First, we imposed a merger and disruption criterion based on the Roche limit and Hill sphere. 
Any binary with a pericenter or apocenter exceeding the Roche limit or lying outside its Hill sphere was excluded, as such systems would either merge or become unbound \citep{2026A&A...706A..63S}.
Only binaries that are Roche- and Hill-stable from the outset were considered.

Second, we added to the observational $K$-limit sample the recent detection of a binary system in the S cluster, known as the D9 binary \citep{2024NatCo..1510608P}.
The age of this system, \(\sim3\,\mathrm{Myr}\) \citep{2024NatCo..1510608P}, is comparable to that of the early-type S-cluster stars, which are \(\lesssim15\,\mathrm{Myr}\) old, with a mean age of \(\sim6\,\mathrm{Myr}\) \citep{2005ApJ...628..246E, 2013ApJ...764..155L, 2017ApJ...847..120H}. Late-type stars were excluded because their older ages, \(\sim3\)--\(10\,\mathrm{Gyr}\) \citep{2019ApJ...872L..15H, 2024PJAB..100...86N}, make them unrepresentative of the young S-star population.
We excluded the remaining G sources because their reported radial velocities \citep[e.g.][]{2020Natur.577..337C, 2025A&A...704A...7P, 2026A&A...707A..79G} have not yet been translated into binary \(K\)-amplitude constraints or calibrated binary-detectability limits. We used Eq.~\eqref{K_amplitude} to compute \(K_{\rm A}\) for the D9 binary, obtaining \(K_{\rm A} \sim 9\,\mathrm{km\,s^{-1}}\), and we incorporated this value into the \(K\) limits listed in Table~7 of \citet{2023ApJ...948...94C}. By combining the $K$ limits of the 16 stars within the S cluster with the $K$ amplitude of the D9 binary, we obtained tighter constraints on the binary fraction and refined its upper limit at a projected distance of $\sim 0.04$~pc. 
In summary, our sample is selected based on the dynamical \(K\)-amplitude information.

Taking the detection of the D9 binary into account, we then derived two constraints on the binary fraction. The upper limit was defined as the fraction of simulations yielding one or fewer detections in 5\% of the runs. The lower limit was defined as the fraction of simulations producing one or more detections in 95\% of the runs.

\begin{figure}[!h]
\centering
\includegraphics[width=0.95\linewidth]{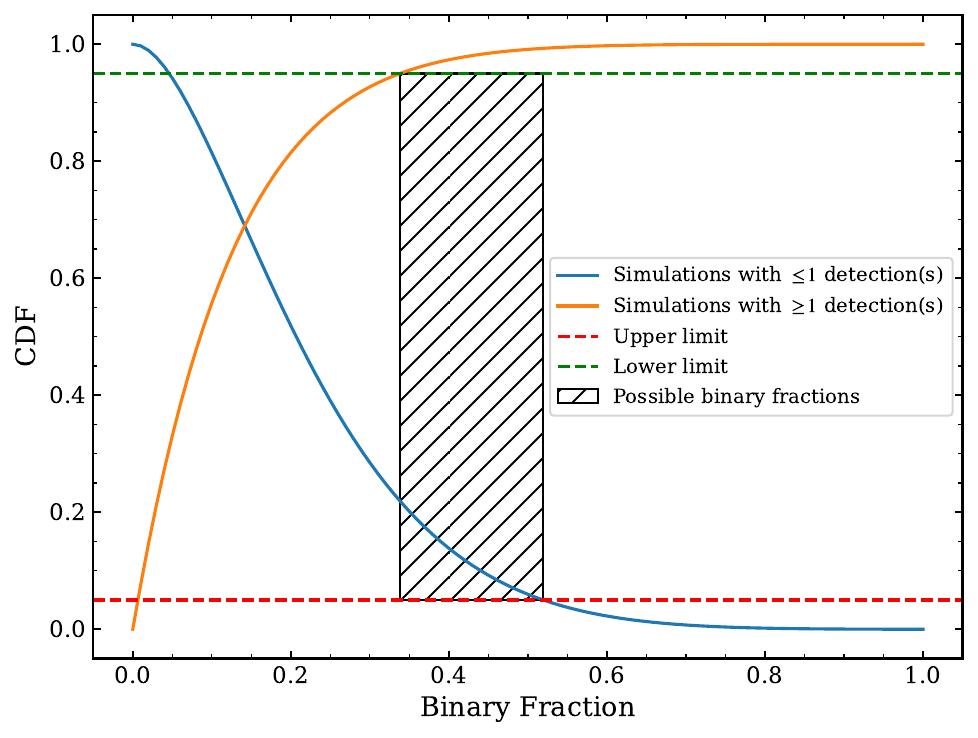}
\caption{CDF of the number of simulations as a function of the binary fraction. The blue line represents the simulations yielding $\leq 1$ detection, and the orange line represents those yielding $\geq 1$ detection. The dashed red and green lines indicate the upper and lower limits, respectively. The dashed black box marks the range of binary fractions that is consistent with both limits.~\label{detections_limits}}
\end{figure}

In Fig.~\ref{detections_limits} we present the cumulative distribution function (CDF) of the number of simulations as a function of the binary fraction. 
Fig.~\ref{detections_limits} shows that the 5\% of the simulations that yield one detection at most correspond to an upper limit of 52\%, while the 95\% of the simulations with one detection at least correspond to a lower limit of 34\% (dashed black box). 
Consequently, the observational binary fraction for the S cluster can be constrained to
\begin{equation}~\label{eq:binary_fraction_obs}
f_{\rm obs} = 0.43 \pm 0.09 \quad (2\sigma) \ .
\end{equation}

\vspace{-2em}
%%%%%%%%%%%%%%%%%%%%%%%%%%%%%%%%%%%%%%%%%%%%%%%%%%%%%%%%%%%%%%%
\section{Theoretical binary fraction}\label{theo_results}
\begin{figure}[!h]
\centering
\includegraphics[width=0.95\linewidth]{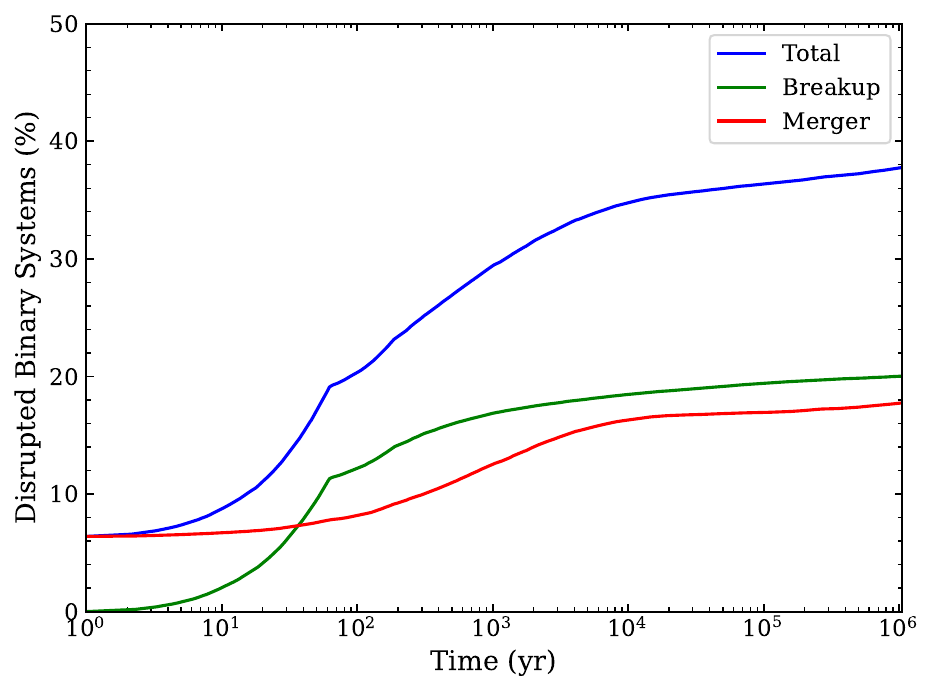}
\caption{Disruption rate of binary systems as a function of time. The red line shows the number of systems that are disrupted through mergers, the green line represents those that are disrupted by gravitational breakup, and the blue line shows the total number of disrupted systems. The break around $60$~yr corresponds approximately to half the orbital period of most binaries in our sample, during which they pass closest to Sgr~A*, where tidal forces are strongest and binary disruption becomes more efficient. The enhanced binary count in this region might be explained by biases in the observational semi-major axis distribution \citep{2017ApJ...837...30G} on which our simulations are based (see methods). 
~\label{bin_frac}}
\end{figure}

We generated a sample of $10^{5}$ binary systems around Sgr~A* following the orbital distributions of the S-cluster stars \citep{2009ApJ...692.1075G, 2017ApJ...837...30G} and massive binary stars \citep{2012Sci...337..444S}, and we evolved them for $10^{6}$~yr using a $N$-body code (see methods).
Three distinct outcomes emerged: 
(i) binaries that were disrupted through tidal interactions with Sgr~A* \citep{1988Natur.331..687H}, 
(ii) binaries that merged as a consequence of the Lidov--Kozai (LK) mechanism \citep{1962P&SS....9..719L, 1962AJ.....67..591K}, and 
(iii) binaries that remained intact throughout the entire simulation. 
We classified these outcomes as breakups, mergers, and survivors, respectively.
Figure~\ref{bin_frac} shows the disruption rate of binary systems as a function of time, distinguishing between breakups, mergers, and the total number of disruptions. In terms of relative frequencies, breakups occur in approximately 20\% of cases, mergers in 18\%, and the remaining 62\% correspond to surviving systems.
\footnote{
    We note that the initial \(7\%\) of mergers follows directly from applying the Roche-limit criterion to binaries sampled from our adopted inner-binary distributions \citep{2012Sci...337..444S}. Resampling binaries initially inside their Roche limit, and retaining only stable systems, does not affect the inferred binary fraction or our conclusions. We did not do so in order to remain faithful to the distributions of \citet{2012Sci...337..444S}.
}

\citet{2012Sci...337..444S} reported that massive stars in the Galactic field, such as those in the S cluster, have a binary fraction of $f_{\rm \infty} = 0.69 \pm 0.09$. 
We assumed for the sake of argument that the S stars were indeed formed in situ. 
For this purpose, we considered a sample of $100$ stars. Following \citep{2012Sci...337..444S}, this implies $69 \pm 9$ binaries and $31 \mp 9$ single stars; for simplicity, we neglect the uncertainties hereafter. 
After evolving the $69$ binaries for $10^{6}$~yr, we found that $62\%$ survive, $20\%$ are disrupted, and $18\%$ merge. In other words, after evolution, we have $43$ surviving binaries, $28$ single stars produced by disruptions, and $12$ single stars resulting from mergers.
Although hypervelocity stars might also form, their number is negligible in our simulations and was therefore omitted from this analysis. 
Consequently, at the end of the evolution, we have $43$ binaries and $71$ single stars, corresponding to a theoretical binary fraction of  
\begin{equation}\label{eq:binary_fraction_theo}
f_{\rm theo} = 0.38 \pm 0.10 \quad (2\sigma) \ .
\end{equation}
\vspace{-2em}
%%%%%%%%%%%%%%%%%%%%%%%%%%%%%%%%%%%%%%%%%%%%%%%%%%%%%%%%%%%%%%%
\section{Spatial distribution evolution of the binary fraction}~\label{binary_frac_pred}
\begin{figure}[!h]
\centering
    \includegraphics[width=0.95\linewidth]{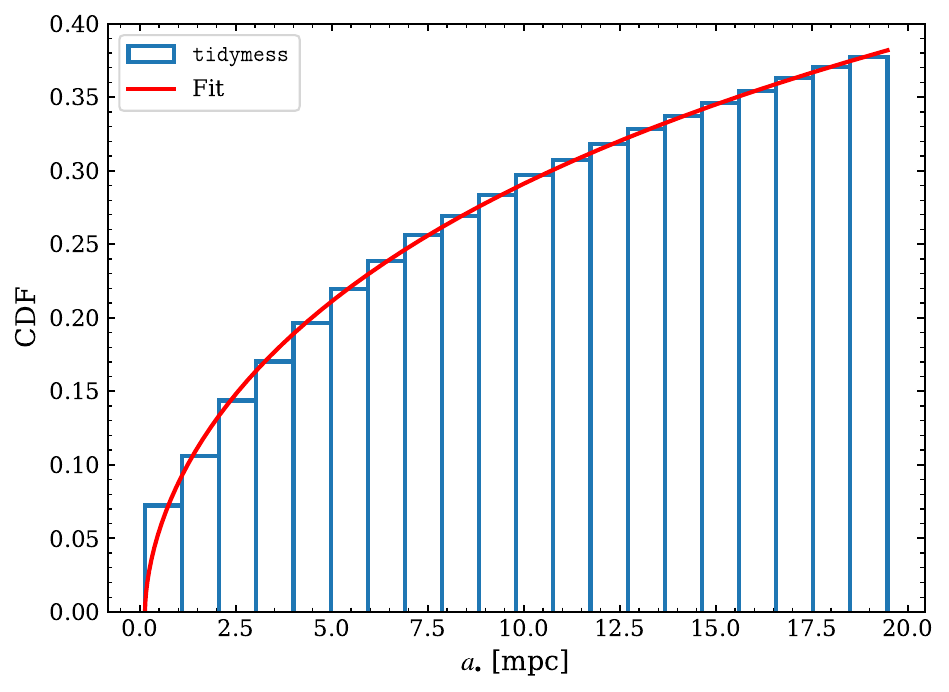}
\caption{Distribution of the binary fraction obtained by the $N$-body code (blue histogram) as a function of the distance to Sgr A* ($a_{\bullet}$). The red line denotes the best fit using Eq.~\eqref{best_fit}. \label{bin_frac_evolution}}
\end{figure}

From the theoretical results, we also inferred how the binary fraction distribution evolves with increasing distance from Sgr~A* (Fig.~\ref{bin_frac_evolution}). 
As expected, near Sgr~A*, the binary fraction is the lowest due to the strong tidal forces in close proximity. 
As the distance to Sgr~A* increases, the binary fraction correspondingly increases.

The binary fraction shown in Fig.~\ref{bin_frac_evolution} follows an exponential dependence on the semi-major axis of the binaries around the supermassive black hole, $a_{\bullet}$. 
To describe this trend, we modelled this evolution using a cumulative Weibull distribution function,
\begin{align}~\label{best_fit}
    &f_{\rm bin}(a_{\bullet}) = f_{\rm \infty} \left(1 - \exp\left[-\left(\frac{a_{\bullet} - r_{\rm t}}{\lambda}\right)^{k} \right] \right),
\end{align}
with the best-fit parameters
\begin{equation}
k =  0.57 \pm 0.02
\quad \text{and} \quad 
\lambda = 0.028 \pm 0.001 \; \mathrm{pc}  \ ,
\end{equation}
where the uncertainties correspond to the $2\sigma$ confidence level. The term $f_{\rm \infty}$ denotes the binary fraction observed in the field for massive binaries \citep{2012Sci...337..444S}, while $r_{\rm t} \approx 10^{-4}$~pc represents the tidal limit of Sgr A* \citep{2005ApJ...631L.117M, 2009MNRAS.392L..31S}.

\begin{figure}[!h]
\centering
\includegraphics[width=0.95\linewidth]{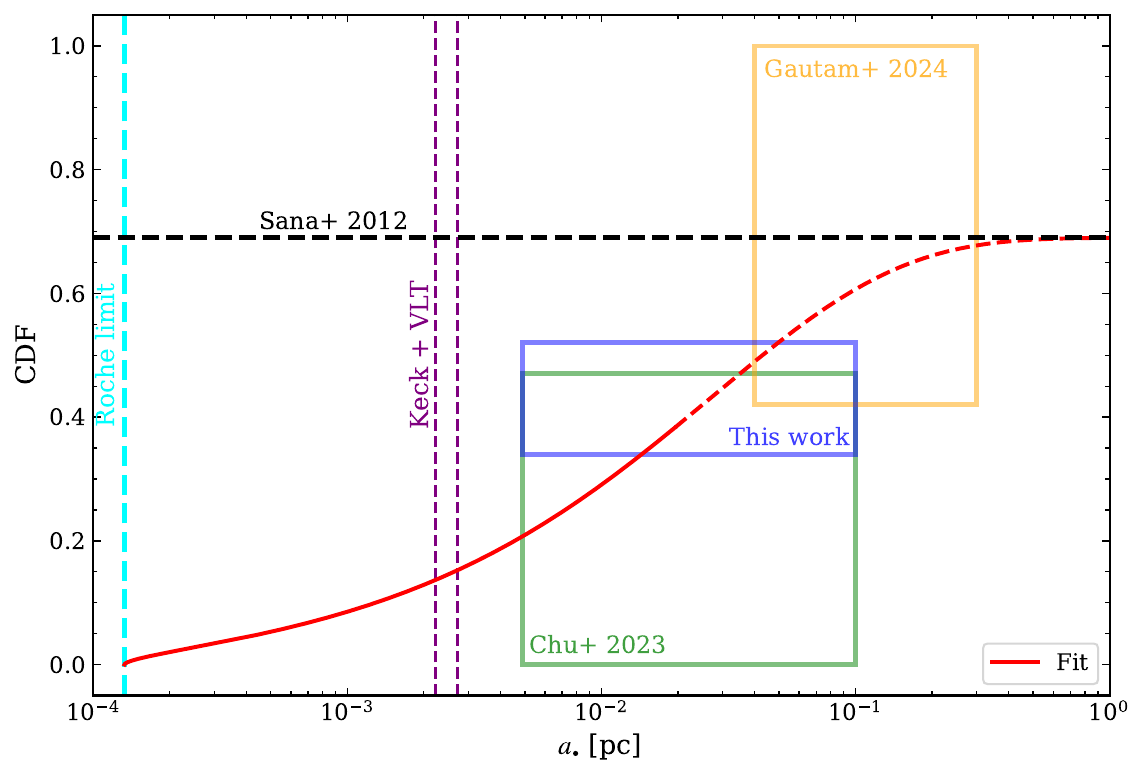}
\caption{Distribution of the binary fraction (red line), including its extrapolation to distances beyond those covered by the $N$-body simulations (dashed red line). The green box indicates the upper limit reported by \citet{2023ApJ...948...94C}. The blue box shows the upper and lower limits we derived. The orange box indicates the lower limit from \citet{2024ApJ...964..164G}. The dashed purple line shows the linear size resolution limit of the Keck Observatory and the VLT. The dashed black line denotes the Galactic field binary fraction \citep{2012Sci...337..444S}. The cyan line gives the tidal limit of the binary-Sgr A* system.
\label{bin_frac_evolution_2}}
\end{figure}

Adopting Eq.~\eqref{best_fit}, we extended $f_{\rm bin}$ to distances reaching into the central parsec of the Galactic centre and thereby overlapping with the ranges considered by \citet{2023ApJ...948...94C} and \citet{2024ApJ...964..164G}. 
Fig.~\ref{bin_frac_evolution_2} shows that the binary fraction estimate (Eq.~\ref{best_fit}) remains below the upper limit reported by \citet{2023ApJ...948...94C} at the corresponding distance ($\sim 0.02$~pc). 
At $\sim 0.04$~pc, using the observational upper and lower limits we derived here, the binary fraction falls within these bounds, demonstrating consistency with the available constraints. 
Likewise, at the distance considered by \citet{2024ApJ...964..164G} ($\sim 0.4$~pc), the binary fraction is also consistent with their lower limit. 
Taken together, these findings support a common formation mechanism for the stellar population, with the observed decline in binary fraction primarily driven by the increasingly strong tidal field in the vicinity of Sgr~A*. 

%%%%%%%%%%%%%%%%%%%%%%%%%%%%%%%%%%%%%%%%%%%%%%%%%%%%%%%%%%%%%%%
\section{Conclusions}
We combined observational and theoretical approaches to constrain the binary fraction of the S cluster and to assess the viability of an in situ formation scenario. 
Reproducing the Monte Carlo method of \citet{2023ApJ...948...94C} and incorporating the newly discovered D9 binary \citep{2024NatCo..1510608P}, we derived an updated observational limit on the S-cluster binary fraction, 
$f_{\rm obs} = 0.43 \pm 0.09$.
From the theoretical side, we evolved $10^{5}$ binary systems for $10^{6}$~yr using an $N$-body code, finding that $20\%$ are disrupted, $18\%$ merge, and $62\%$ survive.
An in situ formation with an initial binary fraction of $69\% \pm 9\%$ \citep{2012Sci...337..444S} yields a final fraction of $f_{\rm theo} = 0.38 \pm 0.10$, which is consistent with current observational constraints in the central parsec. It is worth noting that the binary fraction for B-type stars reported by \citep{2022A&A...658A..69B} is consistent with \citep{2012Sci...337..444S} within $1\sigma$. We adopted \citep{2012Sci...337..444S} because this study is highly complete.

Modelling the radial dependence of the binary fraction with an exponential profile further showed agreement with the observations from $\sim 0.02$ to $\sim 0.4$~pc. This indicates a common formation mechanism throughout this region, with the decline towards Sgr~A* driven primarily by increasing tidal disruption.

Our findings support an in situ formation scenario, such as the gas-shell fragmentation \citep{2018MNRAS.478L.127N}. In our model, massive binaries form where they are observed, without requiring migration or relaxation to reproduce the orbital properties of stars around Sgr~A*. This scenario naturally yields the S-cluster semi-major axis and eccentricity distributions reported by \citet{2017ApJ...837...30G}, making the presence of binaries among these mostly massive stars a natural expectation. In this framework, in situ star formation produces O- and B-type stars in the immediate vicinity of Sgr~A*, some of which might later evolve into stellar mass black holes. Over time, O-type stars are efficiently depleted through direct collisions with the black hole population within \(\sim 5\)~Myr, whereas B-type stars can survive for up to \(\sim 55\)~Myr \citep{2025A&A...695L..19H}. This naturally explains why the currently observed S-cluster population is dominated by B-type stars.

Future improvements in spectroscopic precision and astrometric monitoring will enable us to place tighter constraints on the binary fraction, providing deeper insight into the formation and evolutionary history of the S cluster and its connection to the broader environment of the Galactic centre.

%%%%%%%%%%%%%%%%%%%%%%%%%%%%%%%%%%%%%%%%%%%%%%%%%%%%%%%%%%%%%%%
\begin{acknowledgements}
We thank the anonymous reviewer for several suggestions that improved the manuscript. We also gratefully acknowledge Nuno Morujão for support with computational resources, as well as Devin Chu for valuable discussions and helpful comments.
This work was financed through national funds by FCT - Funda\c{c}\~ao para a Ci\^encia e a Tecnologia, I.P., Portugal, in the framework of the projects
 \href{https://doi.org/10.54499/2024.01252.BD}{2024.01252.BD}, CFisUC UID/04564/2025 (with DOI identifier \href{https://doi.org/10.54499/UID/04564/2025}{10.54499/UID/04564/2025}), and to the Center for Astrophysics and Gravitation (CENTRA/IST/ULisboa) through grant No. UID/PRR/00099/2025 (\href{https://doi.org/10.54499/UID/PRR/00099/2025}{https://doi.org/10.54499/UID/PRR/00099/2025}) and grant No. UID/00099/2025 (\href{https://doi.org/10.54499/UID/00099/2025}{https://doi.org/10.54499/UID/00099/2025}).
TB is supported by the European Union’s Horizon Europe research and innovation
programme under the Marie Skłodowska–Curie grant agreement No 101153423.
We acknowledge the Laboratory for Advanced Computing at the University of Coimbra (\href{https://www.uc.pt/lca}{https://www.uc.pt/lca}) and Rede Nacional de Computação Avançada, under grants 2025.00007.HPCVLAB.UPORTO and 2025.08956.CPCA.A1, for providing the resources to perform the numerical simulations.
\end{acknowledgements}
\vspace{-2em}
%%%%%%%%%%%%%%%%%%%%%%%%%%%%%%%%%%%%%%%%%%%%%%%%%%%%%%%%%%%%%%%
\bibliographystyle{aa} % style aa.bst
\bibliography{aa} % your references Yourfile.bib

%%%%%%%%%%%%%%%%%%%%%%%%%%%%%%%%%%%%%%%%%%%%%%%%%%%%%%%%%%%%%%%
\begin{appendix}
\section{Data availability}
The data used in this study to perform the simulations with the \(N\)-body code \texttt{TIDYMESS} \citep{2023MNRAS.522.2885B}, along with all data required to reproduce the figures and results presented in this work, are publicly available on Zenodo \footnote{\href{https://doi.org/10.5281/zenodo.22917117}{https://doi.org/10.5281/zenodo.22917117}} under the Creative Commons Attribution 4.0 International (CC BY 4.0) license.

The database is designed for use with SQL database engines, such as PostgreSQL and MySQL, as well as Python-based interfaces such as SQLAlchemy. For each binary system considered, it contains detailed information at every simulation time step, given in years, up to the total integration time for systems that remain bound, or up to the disruption time otherwise. The database also provides the inertial position and velocity vectors of Sgr A* and of the primary and secondary components of each binary. Additional quantities include the binary orbital period, eccentricity, primary mass, mass ratio, and cosine of the mutual inclination.

The database can be queried directly using an SQL engine to reproduce all figures presented in this work, as well as to perform additional analyses of the simulation results.

\section{Methods}~\label{methods}
\vspace{-2em}
\subsection{N-body simulations}~\label{N_body_simulation}
We generate a sample of $10^5$ different initial hierarchical three-body systems composed by a inner binary (two bound stars) and Sgr A*.
We then evolve each system for $10^6$~yr using numerical simulations.
This time length value corresponds to a compromise between the computational running time and the dynamical evolution of the system (Fig.~\ref{bin_frac}).

To perform the simulations, we use the open-source $N$-body code \texttt{tidymess} \citep{2023MNRAS.522.2885B}, which employs a time-symmetric adaptive time-step with first post-Newtonian (PN) corrections, 
allowing accurate treatment of close encounters while conserving the total energy \citep{2023MNRAS.519.3281B}. 
Many binary systems in our simulations become highly eccentric, chaotic, or disrupted, necessitating a variable time-step to accurately track their orbital evolution.

\subsubsection{Outer orbits around Sgr~A*}
\begin{figure}[!h]
    \centering
    \makebox[\linewidth]{%
        \includegraphics[width=0.95\linewidth]{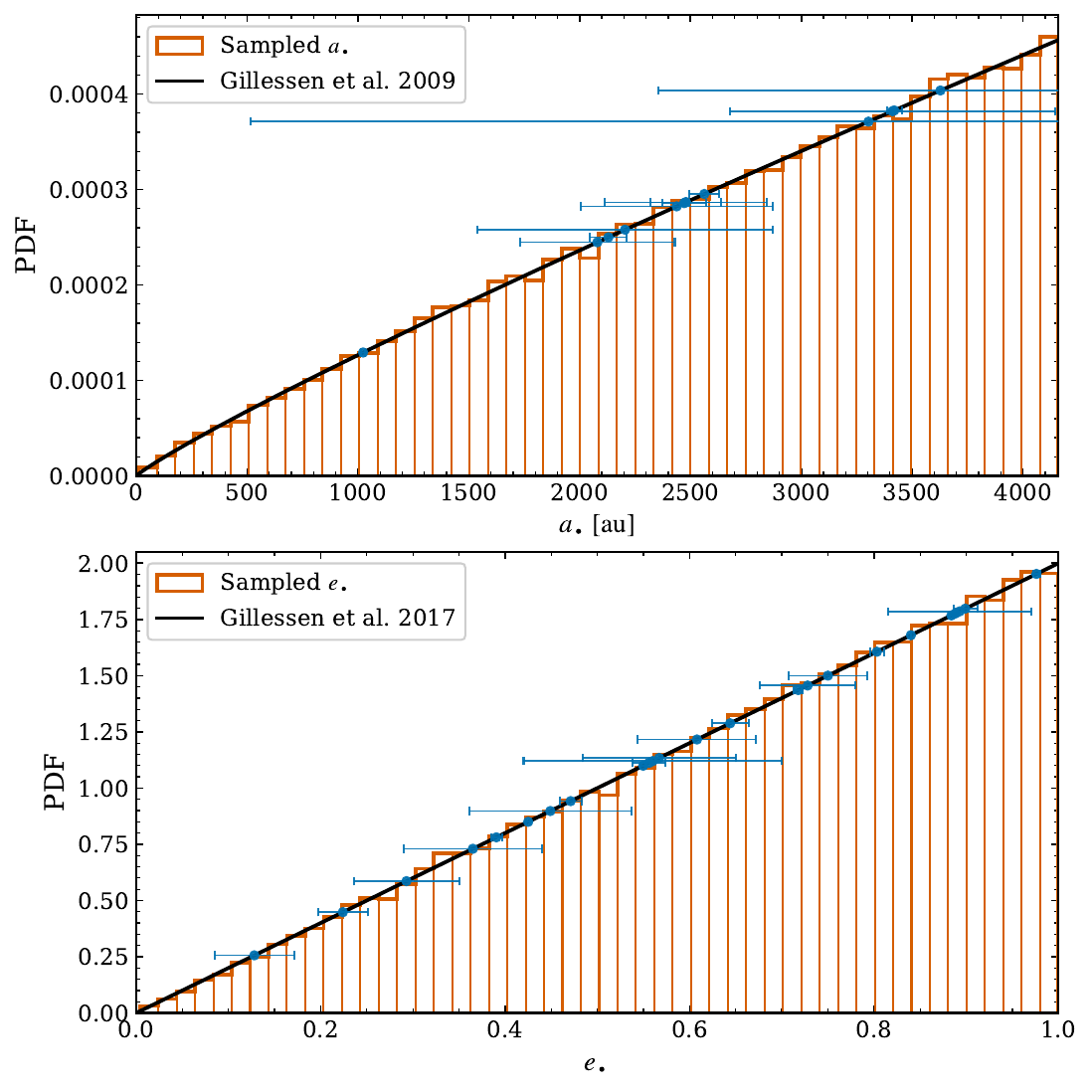}%
    }
    \caption{Distribution of the early-type semi-major axis and eccentricity of the binary orbit centre-of-mass around Sgr A* \citep{2009ApJ...692.1075G, 2017ApJ...837...30G}. The samples shown in orange are generated using the distributions from \cite{2009ApJ...692.1075G, 2017ApJ...837...30G} via the inverse transform sampling method. The black curves represent the corresponding analytical distributions from \cite{2009ApJ...692.1075G, 2017ApJ...837...30G}.}
    \label{Gillessen_distr}
\end{figure}

The orbital elements of the binary orbit centre-of-mass around Sgr* are generated following the early-type distributions of \cite{2009ApJ...692.1075G, 2017ApJ...837...30G} (see our Fig.~\ref{Gillessen_distr}).
Specifically, the initial semi-major axis is drawn from  
\begin{equation}\label{semi_major_axis_outer}  
    f_{\mathrm{p}}\left( a_{\bullet} \right) \propto a_{\bullet}^{0.9} \ ,  
    \quad r_{t} \leq a_{\bullet} \leq 4016 \, \mathrm{au} \ ,
\end{equation}  
where the upper limit corresponds to the extent of the S cluster and the lower limit is given by the tidal radius of Sgr A* \citep{2005ApJ...631L.117M, 2009MNRAS.392L..31S}.

Similarly, the eccentricity is sampled from \citet[Table 3]{2017ApJ...837...30G}
\begin{equation}\label{eccentricity_outer}
    f_{\mathrm{p}}\left( e_{\bullet} \right) \propto e_{\bullet} \ ,  
    \quad 0.0 \leq e_{\bullet} \leq 1.0 \ .
\end{equation}  

\noindent The longitude of the ascending node, $\Omega_{\bullet}$, and argument of pericentre, $\omega_{\bullet}$, are uniformly sampled between $0$ and $2\pi$. 
Finally, the inclination, $I_{\bullet}$, is sampled by uniformly selecting $\cos I_{\bullet}$ in the range $[-1,1]$. 
The mean anomaly, $M_{\bullet}$, is always set at $\pi$, to start the system at the apocentre, where the gravitational interactions are weaker.

\subsubsection{Inner orbits of the binary systems}

For the inner binary system, we must sample both its orbital parameters and the physical properties (masses and radii) of its two stellar components.

To determine the mass distribution of the primary star, $m_{\rm A}$, we use Table~1 from \citet{2018CoTPh..70..735C}, where masses were derived from the K-band apparent magnitudes of stars monitored by \citet{2017ApJ...837...30G}. 
We find the best-fit exponent to be $\alpha = -1.75 \pm 0.03$, consistent with the $\alpha = -1.7 \pm 0.2$ suggested by \citealt{2013ApJ...764..155L} and \citealt{2024A&A...689A.190G}. 
Consequently, we adopt a top-heavy initial mass function (IMF) for our analysis,
\begin{equation}  
    f_{\mathrm{p}}\left( m_{\rm A} \right) \propto m_{\rm A}^{-1.75} \ ,  
    \quad 3.0  \leq m_{\rm A} \leq 15 \, M_{\odot} \ ,
\end{equation} 
where the minimum and maximum values are chosen based on spectroscopic mass measurements \citep{2017ApJ...847..120H} and K-band magnitudes of observed S stars \citep{2018CoTPh..70..735C}.
Given the primary mass, its radius is determined using the zero-age main-sequence (ZAMS) mass-radius relation \citep{1991Ap&SS.181..313D}. 

\begin{figure}[!h]
    \centering
    \makebox[\linewidth]{%
        \includegraphics[width=0.95\linewidth]{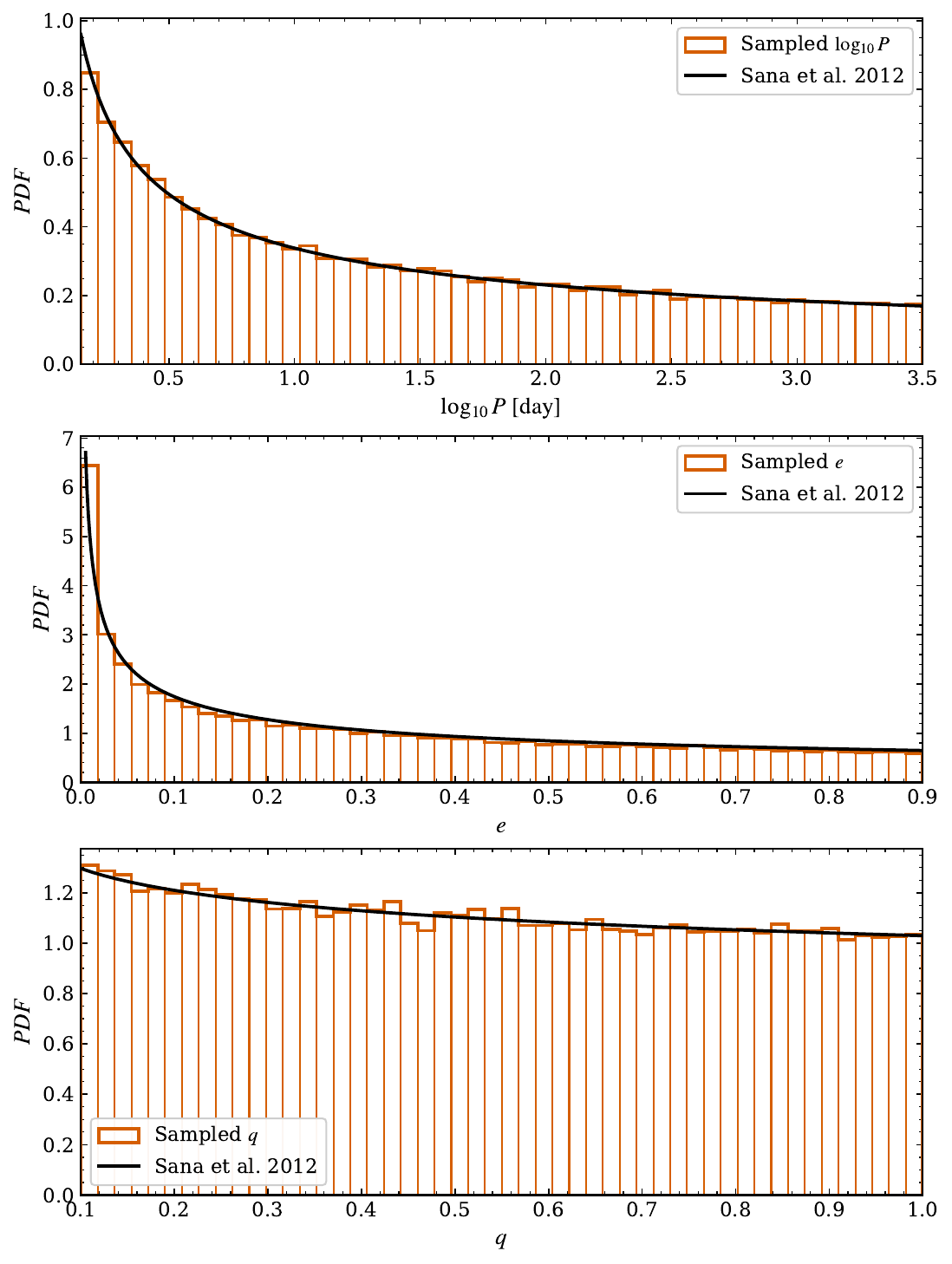}%
    }
    \caption{Distribution of the inner binary orbital period, eccentricity, and mass ratio following \citep{2012Sci...337..444S}. The samples shown in orange are generated using the distributions from \cite{2012Sci...337..444S} via the inverse transform sampling method. The black curves represent the corresponding analytical distributions from \cite{2012Sci...337..444S}.}
    \label{Sana_distr}
\end{figure}

To determine the secondary star’s mass and radius, we first draw the mass ratio, $q = m_B / m_A$, from the distribution \citep{2012Sci...337..444S} (see our Fig.~\ref{Sana_distr})
\begin{equation} \label{ratio_inner}  
    f_{\mathrm{p}}(q) \propto q^{-0.1},  
    \quad 0.1 \leq q \leq 1.0,
\end{equation}  
Once $q$ is determined, the secondary mass is calculated as $m_B = q m_A$, and its radius is again obtained using the ZAMS mass-radius relation.

The inner binary's orbital period is drawn from the distribution \citep{2012Sci...337..444S} (see our Fig.~\ref{Sana_distr})
\begin{equation}\label{period_inner}   
    f_{\mathrm{p}}\left( \log_{10} P \right) \propto \log_{10} P^{-0.55} \ ,  
    \quad 10^{0.15} \leq P \leq 10^{3.5} \, \mathrm{day} \ ,
\end{equation}  
and its eccentricity from \cite{2012Sci...337..444S}  
\begin{equation}\label{ecc_inner}  
    f_{\mathrm{p}}(e) \propto e^{-0.45} \ ,  
    \quad 0.0 \leq e \leq 0.9 \ .
\end{equation}  

The longitude of the ascending node, $\Omega$, argument of pericentre, $\omega$, and mean anomaly, $M$, are uniformly sampled between $0$ and $2\pi$, while the inclination, $I$, is sampled by uniformly selecting $\cos I$ in the range $[-1,1]$.

\section{Lidov--Kozai timescale}~\label{app:KL_timescale}
The Lidov--Kozai mechanism is a secular perturbation in which the inner binary periodically exchanges mutual inclination and eccentricity, facilitating unstable orbits and mergers \citep[e.g.][]{1962P&SS....9..719L, 1962AJ.....67..591K}. Its characteristic timescale is \citep[e.g.][]{2015MNRAS.451.1341L, 2016ARA&A..54..441N}
\begin{align}
    t_{\rm LK} \sim 
    \frac{2\pi a_{\bullet}^{3}\left(1 - e_{\bullet}^{2} \right)^{3/2}
    \sqrt{\left(m_{\rm A} + m_{\rm B}\right)\left(1 - e^{2}\right)}}
    {\sqrt{G}a^{3/2}m_{\bullet}} \, ,
    \label{t_quad}
\end{align}
where \(a_{\bullet}\) and \(e_{\bullet}\) are the semi-major axis and eccentricity of the outer orbit, \(a\) and \(e\) those of the inner orbit, and \(m_{\rm A}\), \(m_{\rm B}\), and \(m_{\bullet}\) are the masses of the primary, the secondary, and the central super massive black hole, respectively.

\begin{figure}
    \centering
    \includegraphics[width=0.95\linewidth]{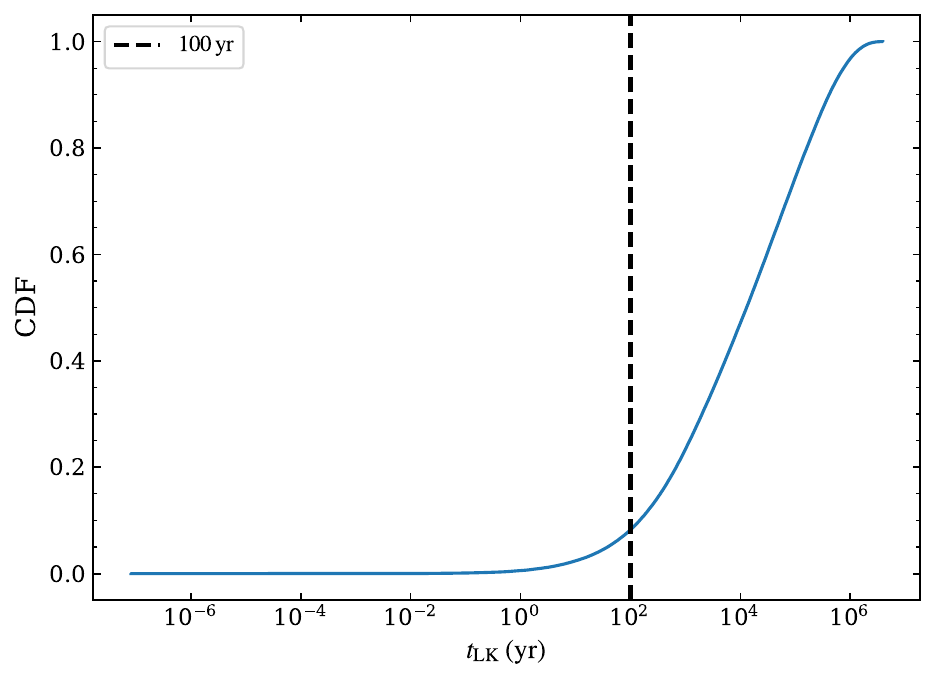}
    \caption{CDF of the initial binaries as a function of the Lidov--Kozai timescale, \(t_{\rm LK}\). The blue line shows the CDF, while the dashed black  line marks \(t_{\rm LK}=10^{2}\,\mathrm{yr}\).}
    \label{fig:hist_t_LK}
\end{figure}

In Fig.~\ref{fig:hist_t_LK}, we show the cumulative distribution function (CDF) of initial binaries as a function of the Lidov--Kozai timescale. Only \(8\%\) of the binaries have \(t_{\rm LK} \leq 10^{2}\,\mathrm{yr}\), while \(92\%\) have longer Lidov--Kozai timescales. Since the Lidov--Kozai mechanism can drive unstable orbits and induce mergers, the larger number of binaries with \(t_{\rm LK} \gtrsim 10^{2}\,\mathrm{yr}\) explains why the merger rate begins to rise around \(10^{2}\,\mathrm{yr}\).
\section{Binary evaporation}~\label{app:binary_evaporation}
\citet{2016MNRAS.460.3494S} consider a binary model and study how many binaries would merge or evaporate in the context of the Galactic centre. Binary evaporation occurs when binaries soften through interactions with other stars in the cluster, which can make the binaries unbound. Binaries that soften are usually referred to as soft binaries, although interactions with cluster stars may also harden binaries. Following \citet{2016MNRAS.460.3494S}, we assume that hard binaries, if formed in the Galactic centre, can survive up to \(10\,\mathrm{Gyr}\).

In Fig.~5 of \citet{2016MNRAS.460.3494S}, the number of evaporated binaries starts to increase around \(1\,\mathrm{Myr}\), that is, after the integration time considered in this work. Hence, after \(1\,\mathrm{Myr}\), this effect should be taken into account. The timescale of this effect is given by \citep{2016MNRAS.460.3494S}
\begin{equation}
    t_{\mathrm{EV}}
    =
    \sqrt{\frac{3}{\pi}}\frac{\sigma}
    {32G\rho a \ln \Lambda}
    \frac{m_{\rm A} + m_{\rm B}}{m_{\rm pert}},
\end{equation}
where \(\rho\) is the stellar mass density in the cluster \citep{2010RvMP...82.3121G},
\begin{equation}
    \rho = 1.35 \times 10^6\, M_\odot\, \mathrm{pc}^{-3}
    \left(\frac{a_\bullet}{0.25\,\mathrm{pc}}\right)^{-1},
\end{equation}
\(\ln \Lambda\) is the Coulomb logarithm, with \(\Lambda = 15\), \(\sigma\) is the velocity dispersion of the Galactic centre \citep{2011MNRAS.412..187K},
\begin{equation}
    \sigma = 280\,\mathrm{km\,s^{-1}}\sqrt{0.1\,\mathrm{pc}/a_\bullet},
\end{equation}
and \(m_{\rm pert}\) is the average mass of a cluster star, which we assume \(m_{\rm pert}=1\,M_\odot\), following \citep{2009ApJ...700.1933H, 2016MNRAS.460.3494S}.

\begin{figure}
    \centering
    \includegraphics[width=0.95\linewidth]{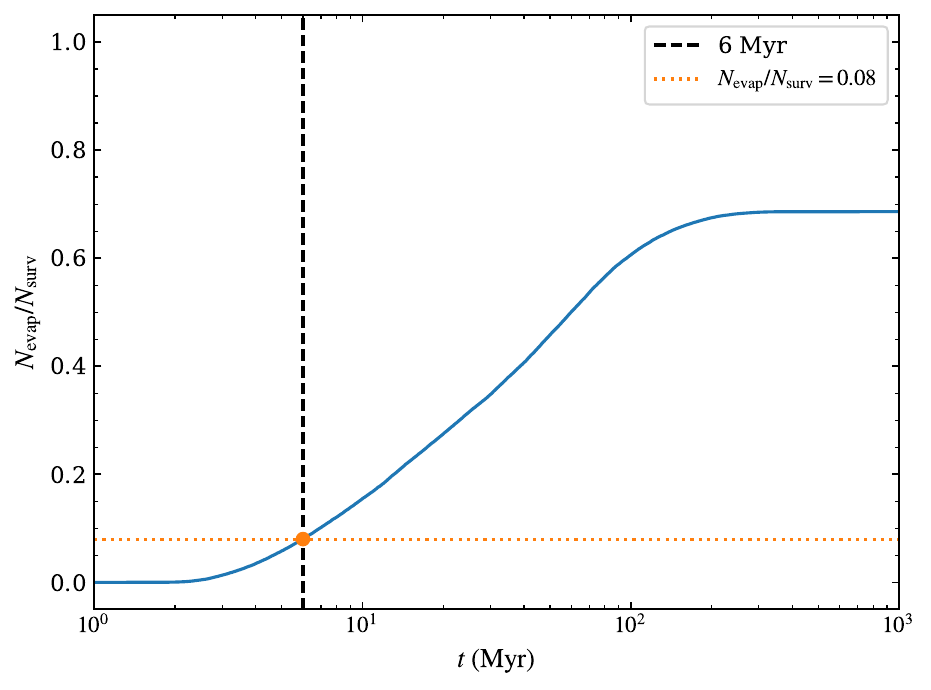}
    \caption{
    Cumulative fraction of soft binaries as a function of time, \(t\), normalised by the total number of binaries surviving at \(1\,\mathrm{Myr}\). The blue curve shows this cumulative fraction. The dotted~orange line marks the fraction with \(t<6\,\mathrm{Myr}\), approximately \(8\%\), while the dashed black line denotes \(t=6\,\mathrm{Myr}\), the estimated age of the young stellar cluster \citep{2013ApJ...764..155L}.
    }
    \label{fig:cumu_frac_t_evap}
\end{figure}

We estimate how many surviving binaries would evaporate by comparing their evaporation timescale, \(t_{\rm evap}\), with the additional evolution time, \(t_{\rm evol}\). Since the young stellar population has an estimated age of \(6\,\mathrm{Myr}\) \citep{2013ApJ...764..155L}, and our integrations stop at \(1\,\mathrm{Myr}\), we set \(t_{\rm evol}=5\,\mathrm{Myr}\). If \(t_{\rm evap}<t_{\rm evol}\), we consider the binary evaporated; otherwise, we consider it stable up to \(6\,\mathrm{Myr}\). In Fig.~\ref{fig:cumu_frac_t_evap}, we show the cumulative fraction of soft binaries, normalised by the total number of binaries surviving at \(1\,\mathrm{Myr}\), as a function of time, \(t\). 
% The dashed~black line marks \(t=6\,\mathrm{Myr}\), while the dotted~orange line marks the corresponding fraction.
%
We see that only \(8\%\) of the surviving binaries would evaporate over the next \(5\,\mathrm{Myr}\). This gives a surviving binary fraction of \(57\%\), instead of the \(62\%\) reported in Sec.~\ref{theo_results}. To infer the corrected theoretical binary fraction at \(6\,\mathrm{Myr}\), we write the binary fraction at \(1\,\mathrm{Myr}\) as
\begin{equation}
    f_{\rm bin} = \frac{N_{\rm B}}{N_{\rm B} + N_{\rm S}},
\end{equation}
where \(N_{\rm B}\) is the number of binary systems and \(N_{\rm S}\) is the number of single systems. After 5~Myr, the number of binary systems decreases due to evaporation, so
\begin{equation}
    N^{\prime}_{\rm B} = \left(1 - \epsilon\right)N_{\rm B}, 
    \quad 
    N^\prime_{\rm S} = N_{\rm S} + 2 \epsilon N_{\rm B},
\end{equation}
where \(\epsilon\) is the evaporation factor. In our case, \(\epsilon = 0.08\). The binary fraction at 6~Myr may then be written as
\begin{equation}
    f^{\prime}_{\rm bin} 
    = \frac{N^{\prime}_{\rm B}}{N^{\prime}_{\rm B} + N^{\prime}_{\rm S}} 
    = \frac{\left(1 - \epsilon\right)f_{\rm bin}}{1 +  \epsilon f_{\rm bin}}
    = 0.34 \pm 0.09 \, (2 \sigma),
\end{equation}
which is still in line with the observational constraint on the binary fraction in Eq.~\eqref{eq:binary_fraction_obs} and with \(f_{\rm theo}\) in Eq.~\eqref{eq:binary_fraction_theo}.

\end{appendix}

\end{document}